\documentstyle[preprint,aps,epsf]{revtex}

\newcommand {\bea}{\begin{eqnarray}}
\newcommand {\eea}{\end{eqnarray}}
\newcommand {\be}{\begin{equation}}
\newcommand {\ee}{\end{equation}}

\newcommand {\Dslash}{D\!\!\!/}

\begin{document}

\preprint{SUNY-NTG-01-41}

\title{Mass Terms in Effective Theories of High Density 
Quark Matter}

\author{T.~Sch\"afer$^{1,2}$}

\address{
$^1$Department of Physics, SUNY Stony Brook,
Stony Brook, NY 11794\\ 
$^2$Riken-BNL Research Center, Brookhaven National 
Laboratory, Upton, NY 11973}

\maketitle

\begin{abstract}
  We study the structure of mass terms in the effective theory
for quasi-particles in QCD at high baryon density. To next-to-leading
order in the $1/p_F$ expansion we find two types of mass terms, 
chirality conserving two-fermion operators and chirality violating
four-fermion operators. In the effective chiral theory for Goldstone
modes in the color-flavor-locked (CFL) phase the former terms 
correspond to effective chemical potentials, while the latter 
lead to Lorentz invariant mass terms. We compute the masses 
of Goldstone bosons in the CFL phase, confirming earlier results
by Son and Stephanov as well as Bedaque and Sch\"afer. We show
that to leading order in the coupling constant $g$ there is no 
anti-particle gap contribution to the mass of Goldstone modes, and 
that our results are independent of the choice of gauge. 
 
\end{abstract}

\newpage

\section{Introduction}
\label{sec_intro}

 Quark matter at high baryon density exhibits a rich phase structure
\cite{Bailin:1984bm,Alford:1998zt,Rapp:1998zu,Rajagopal:2000wf,Alford:2001dt,Schafer:2000et}, 
reminiscent of the many phases encountered in ordinary condensed matter 
systems. A particularly symmetric phase is the color-flavor-locked (CFL) 
phase of three flavor quark matter \cite{Alford:1999mk}. This phase is 
believed to be the true ground state of ordinary matter at very large 
density \cite{Schafer:1999ef,Schafer:1999fe,Evans:2000at}. There 
are also speculations that CFL matter may exist in the core of 
neutron stars. At less-than-asymptotic densities relevant to 
astrophysical objects distortions of the pure CFL state due to 
non-zero quark masses are likely to be important
\cite{Alford:1999pa,Schafer:1999pb,Alford:2001ze,Schafer:2000ew,Bedaque:2001je,Alford:2001zr,Kaplan:2001qk}.
In the present work we wish to study this problem through the 
use of two effective field theories. 

 The first of these, CFL chiral theory \cite{Casalbuoni:1999wu},
is the chiral effective theory that governs the behavior of 
collective states with excitation energies smaller than the gap. 
Understanding the structure of mass terms in this theory allows
us to determine the true ground state \cite{Bedaque:2001je,Kaplan:2001qk}
and the spectrum of pseudo-Goldstone bosons 
\cite{Son:1999cm,Rho:2000xf,Hong:2000ei,Manuel:2000wm,Rho:2000ww,Beane:2000ms,Hong:2000ng}
at finite quark mass. The second effective theory, high density 
effective theory \cite{Hong:2000tn,Hong:2000ru}, governs the 
interaction of quasi-particles and holes with excitation energies
less than the Fermi energy in QCD at high baryon density. This 
theory simplifies perturbative QCD calculations in the limit where 
the Fermi momentum is very large, $p_F\to\infty$. Understanding 
the structure of mass corrections in this theory allows us to perform 
efficient calculations of mass terms in the CFL chiral theory. These 
calculations are based on the idea of matching. Matching expresses 
the requirement that Greens functions in the effective theories 
below and above the scale set by gap agree. In the present work, 
we will focus on a particularly simple quantity, the mass 
dependence of the vacuum energy.

 This paper is organized as follows. In section \ref{sec_hdet}
we introduce the effective theory for quasi-particles and holes
in high density QCD and in section \ref{sec_hdetm} we study
the mass terms in this theory. In sections \ref{sec_agap} and 
\ref{sec_ds_agap} we study the possible role of anti-particle 
gap parameters. In section \ref{sec_CFLchi} we discuss the 
matching calculation that determines the masses of Goldstone
bosons in the CFL chiral theory.

\section{High Density Effective Theory (HDET)}
\label{sec_hdet}

 The QCD Lagrangian in the presence of a chemical potential
is given by
\be
\label{qcd}
 {\cal L} = \bar\psi \left( i\Dslash +\mu\gamma_0 \right)\psi
 -\bar\psi_L M\psi_R - \bar\psi_R M^\dagger \psi_L 
 -\frac{1}{4}G^a_{\mu\nu}G^a_{\mu\nu},
\ee
where $M$ is a complex quark mass matrix which transforms as
$M\to LMR^\dagger$ under chiral transformations $(L,R)\in
SU(3)_L\times SU(3)_R$ and $\mu$ is the baryon chemical potential.
In the vicinity of the Fermi surface the relevant degrees of
freedom are particle and hole excitations which move with the
Fermi velocity $v_F$. We shall describe these excitations in 
terms of the field $\psi_+(\vec{v}_F,x)$. At tree level, the 
quark field $\psi$ can be decomposed as $\psi=\psi_++\psi_-$ 
where $\psi_\pm=\frac{1}{2}(1\pm\vec{\alpha}\cdot\hat{v}_F)\psi$. 
Integrating out the $\psi_-$ field at leading order in $1/p_F$ 
we get \cite{Hong:2000tn,Hong:2000ru,Casalbuoni:2001na,Beane:2000ms}
\bea
\label{fs_eff}
{\cal L} &=& 
 \psi_{L+}^\dagger (iv\cdot D) \psi_{L+}
  - \frac{ \Delta}{2}\left(\psi_{L+}^{ai} C \psi_{L+}^{bj}
 \left(\delta_{ai}\delta_{bj}-
           \delta_{aj}\delta_{bi} \right) 
           + {\rm h.c.} \right) \nonumber \\ 
& & \hspace{0.5cm}\mbox{}
  - \frac{1}{2p_F} \psi_{L+}^\dagger \left(  (\Dslash_\perp)^2 
  + MM^\dagger \right)  \psi_{L+}  
  + \left( R\leftrightarrow L, M\leftrightarrow M^\dagger \right)  + \ldots ,
\eea
where $D_\mu=\partial_\mu+igA_\mu$, $v_\mu=(1,\vec{v})$ and
 $i,j,\ldots$ and $a,b,\ldots$ denote flavor and color 
indices. The longitudinal and transverse components of a vector $B_\mu$ 
are defined by $(B_0,\vec{B})_{\|}=(B_0,\vec{v}(\vec{B}\cdot 
\vec{v}))$ and $(B_\mu)_\perp = B_\mu-(B_\mu)_{\|}$. In order to perform 
perturbative calculations in the superconducting phase we have added a 
tree level gap term $\psi_{L,R} C\Delta \psi_{L,R}$. For consistency
we have to subtract this term from the interacting part of the
Lagrangian, ${\cal L}_{int}=-\psi_{L,R} C\Delta \psi_{L,R}$. The 
magnitude of $\Delta$ is determined order by order in perturbation 
theory from the requirement that the free energy is stationary 
with respect to $\Delta$.

\section{Mass contribution to the vacuum energy}
\label{sec_hdetm}

 We would like to compute the shift in vacuum energy due to 
non-zero quark masses and match the result against the shift
computed in the effective theory for the Goldstone bosons. For 
this purpose we need to determine mass corrections to the high 
density effective theory. At $O(1/p_F)$ there is only one 
operator,
\be 
\label{kin}
{\cal L} = -\frac{1}{2p_F} \left( \psi_{L+}^\dagger MM^\dagger \psi_{L+}
 + \psi_{R+}^\dagger M^\dagger M\psi_{R+} \right).
\ee
This term arises from expanding the kinetic energy of a massive
fermion around $p=p_F$. At $O(1/p_F)$ there is no mass correction 
to the quark-gluon vertex. Indeed, integrating out the $\psi_-$ 
field we get
\be
\label{q+AMq}
{\cal L} = \frac{g}{2p_F} \psi^\dagger_{+,L} \left[
 \gamma_0 M \vec{\alpha}\cdot\vec{A}_\perp 
+\vec{\alpha}\cdot\vec{A}_\perp \gamma_0 M\right]\psi_{+,R}
 = 0 .
\ee
There is also no mass correction to the emission of an electric
gluon. This is the case because $\psi^\dagger_-(\vec{v}')gA_0 
\psi_+(\vec{v})$ vanishes in the forward direction, $\vec{v}'=
\vec{v}$. There is, however, a mass dependent four-fermion 
contact term induced by electric gluon exchange. One way to
see this is to consider off-forward amplitudes with $\vec{v}'
\neq\vec{v}$, integrate out hard gluon exchanges, and take 
the forward limit $\vec{v}'\to \vec{v}$ in the end. In the 
case $\vec{v}'\neq\vec{v}$ we find, after integrating out 
$\psi_-$ at tree level, a mass correction to the electric 
gluon vertex
\be
\label{el_cor}
{\cal L} = -\frac{g}{p_F}\psi^\dagger_{L,+}(\vec{v}') \gamma_0 MA_0
   \psi_{R,+}(\vec{v}) + 
  \Big(L\leftrightarrow R, M\leftrightarrow M^\dagger \Big).
\ee
This term still vanishes in the limit $\vec{v}'\to\vec{v}$,
but it gives a non-vanishing contribution to quark-quark 
scattering in the forward direction. This is the case 
because the vanishing amplitude in the forward direction 
cancels against the collinear singularity in the gluon 
propagator. 

 We can construct the four-fermion operator by integrating 
out hard gluons with momenta $p\sim p_F$. Because the vertex
(\ref{el_cor}) vanishes in the forward direction the contribution
from soft gluons with $p\sim gp_F$ is suppressed. We find
\bea 
\label{4f1}
{\cal L} &=&
\frac{g^2}{8p_F^2} \left[ 
 \Big(\psi^\dagger_{+,L}(\vec{v}')\gamma_0M\lambda^a\psi_{+,R}(\vec{v})
                              \Big)
 \Big(\psi^\dagger_{+,L}(\vec{v}')\gamma_0M\lambda^a\psi_{+,R}(\vec{v})
                              \Big)  \right. \\
 && \hspace{0.5cm}\mbox{}\left. +
 \Big(\psi^\dagger_{+,L}(\vec{v}')\gamma_0M\lambda^a\psi_{+,R}(\vec{v})
                              \Big)
 \Big(\psi^\dagger_{+,R}(\vec{v})\gamma_0M^\dagger\lambda^a\psi_{+,L}
      (\vec{v}')\Big)
  + \Big(L\leftrightarrow R, M\leftrightarrow M^\dagger\Big)\right]
 \nonumber \\
 & & \hspace{0.5cm}\mbox{} \cdot 
 \frac{1}{2p_F^2(1-\vec{v}'\cdot\vec{v})} \nonumber
\eea
We can make the $\vec{v}'\to\vec{v}$ limit more explicit by	 
Fierz-rearranging the four-fermion terms in (\ref{4f1}). For the 
first term we use 
\bea 
\lefteqn{
 \Big(\psi^\dagger_L(\vec{v}')\gamma_0\psi_R(\vec{v})\Big)
 \Big(\psi^\dagger_L(\vec{v}')\gamma_0\psi_R(\vec{v})\Big) }
\hspace{1cm} && \\
&=&  \frac{1}{2} \left[
 \Big(\psi_L^\dagger(\vec{v}')C\psi_L^\dagger(\vec{v}')\Big)
 \Big(\psi_R(\vec{v})C\psi_R(\vec{v})\Big)
-\Big(\psi_L^\dagger(\vec{v}')\vec{\alpha}C\psi_L^\dagger(\vec{v}')\Big)
 \Big(\psi_R(\vec{v})C\vec{\alpha}\psi_R(\vec{v})\Big)
 \right] \nonumber \\
 &=&  \frac{1}{2} 
 \Big(\psi_L^\dagger C\psi_L^\dagger \Big)
 \Big(\psi_R C\psi_R \Big)
 \left( 1 -\vec{v}\cdot\vec{v}'\right). \nonumber 
\eea
Here we have dropped the subscript ``$+$'' and suppressed the color 
and flavor structure of the fields. A similar identity can be 
derived for the second term in (\ref{4f1}). We observe that the 
factor $1-\vec{v}\cdot\vec{v}'$ cancels and that in the limit 
$\vec{v}'\to\vec{v}$ we are left with a local four-fermion
interaction. Collecting (\ref{kin}) and (\ref{4f1}) we get the 
following result for the mass terms in the high density effective 
theory to $O(1/p_F^4)$
\bea
\label{hdet_m}
 {\cal L} &=& -\frac{1}{2p_F} \left( \psi_{L}^\dagger MM^\dagger \psi_{L}
 + \psi_{R}^\dagger M^\dagger M\psi_{R} \right) \\
 & & \mbox{} + \frac{g^2}{8p_F^4}
 \left( ({\psi^A_L}^\dagger C{\psi^B_L}^\dagger)
        (\psi^C_R C \psi^D_R) \Gamma^{ABCD} +
        ({\psi^A_L}^\dagger \psi^B_L) 
        ({\psi^C_R}^\dagger \psi^D_R) \tilde{\Gamma}^{ACBD} \right)
        \nonumber \\ 
 & & \mbox{} 
 + \Big(L\leftrightarrow R, M\leftrightarrow M^\dagger \Big)
 + \ldots \nonumber . 
\eea
Here, we have rewritten the four-fermion terms using the CFL 
eigenstates $\psi^A$ defined by $\psi^a_i=\psi^A (\lambda^A)_{ai}
/\sqrt{2}$, $A=0,\ldots,8$ where the Gell-Mann matrices $\lambda^i$
are normalized as ${\rm Tr}[\lambda^i\lambda^j]=2\delta^{ij}$
and $\lambda^0=\sqrt{2/3}$. This is simply a choice of basis, 
it does not imply that (\ref{hdet_m}) is only valid in the CFL phase. 
The tensors $\Gamma$ and $\tilde{\Gamma}$ are defined by
\bea 
 \Gamma^{ABCD} &=& \frac{1}{8}\left\{ {\rm Tr} \left[ 
    \lambda^A M(\lambda^D)^T \lambda^B M (\lambda^C)^T\right]
 -\frac{1}{3} {\rm Tr} \left[
    \lambda^A M(\lambda^D)^T \right]
    {\rm Tr} \left[
    \lambda^B M (\lambda^C)^T\right] \right\},\\
 \tilde{\Gamma}^{ABCD} &=& \frac{1}{8}\left\{ {\rm Tr} \left[ 
    \lambda^A M(\lambda^D)^T \lambda^C M^\dagger (\lambda^B)^T\right]
 -\frac{1}{3} {\rm Tr} \left[
    \lambda^A M(\lambda^D)^T \right]
    {\rm Tr} \left[
    \lambda^C M^\dagger (\lambda^B)^T\right] \right\}\nonumber .
\eea
We should note that equ.~(\ref{hdet_m}) can also be derived 
directly, by computing the chirality violating quark-quark 
scattering amplitude $T((q_R)_i^a+(q_R)_j^b\to (q_L)_k^c+
(q_L)_l^d)$ as well as the mass correction to $T((q_R)_i^a+
(q_L)_j^b\to (q_L)_k^c+(q_R)_l^d)$. The corresponding tree 
level diagrams are shown in Fig.~\ref{fig_4f}. We find that
to leading order in $g$ the scattering amplitudes are independent 
of the scattering angle and that they can be represented by
the contact terms in equ.~(\ref{hdet_m}). We also observe that 
the effective lagrangian (\ref{hdet_m}) is consistent with the 
approximate gauge symmetry of Bedaque and Sch\"afer 
\cite{Bedaque:2001je}. This approximate gauge symmetry reflects 
the fact that to leading order in $1/p_F$ mass terms appear as 
effective chemical potentials $MM^\dagger/(2p_F)$ and $M^\dagger 
M/(2p_F)$ for left and right handed fermions, respectively. 

 We can now compute the shift in the vacuum energy due to
the mass terms in equ.~(\ref{hdet_m}). At $O(M^2)$ and to 
leading order in $1/p_F$ there is a contribution of the form 
${\rm Tr} [MM^\dagger]$ from the two-fermion operators in 
equ.~({\ref{hdet_m}). This contribution is the same in 
the normal and superfluid phases, and it does not 
correspond to a Goldstone boson mass term. This is the case 
because ${\rm Tr}[MM^\dagger]$ cannot be matched against
a term in the CFL chiral theory which contains the chiral
field $\Sigma$. At $O(M^2)$, the only terms in the vacuum 
energy which correspond to Goldstone boson mass terms 
are ${\rm Tr}[M^2]$, $({\rm Tr}[M])^2$ and ${\rm Tr}[M]
{\rm Tr}[M^\dagger]$. 

 At next-to-leading order in $1/p_F$ there is a mass 
correction to the vacuum energy due to the four-fermion 
operators in equ.~(\ref{hdet_m}). This contribution comes 
from the diagrams shown in Fig.~\ref{fig_4fvac}. The first 
diagram is proportional to the square of the superfluid
density
\be 
\label{qq_cond}
\langle \psi^A_L C\psi^B_L\rangle 
 = \delta^{AB} \int\frac{d^4p}{(2\pi)^4} 
    \frac{\Delta_A(p_0)}{p^2-\epsilon_p^2-\Delta_A^2(p_0)}
 =  \delta^{AB}\Delta^A\frac{3\sqrt{2}\pi}{g}  
     \left(\frac{p_F^2}{2\pi^2}\right) ,
\ee
with $\Delta^A=C^A\Delta$ and $C^A=(2,-1)$ for $A=(0,1\ldots 8)$.
In deriving (\ref{qq_cond}) we have used the approximate solution 
of the gap equation $\Delta(p_0) = \Delta(p_0\!=\! 0)\sin(\frac{g}
{3\sqrt{2}\pi}\log(\mu/p_0))$. The color-flavor factor is given
by
\be 
\label{cfl_fac}
 \sum_{A,B=0}^8 C^A C^B \Gamma^{AABB} 
 = -\frac{4}{3} \left\{  \Big( {\rm Tr}[M]\Big)^2 -
        {\rm Tr}\Big[ M^2\Big]   \right\} .
\ee
The second diagram is proportional to ${\rm Tr}[MM^\dagger]$ and 
does not contribute to Goldstone boson masses. We note that there 
is no contribution of the form ${\rm Tr}[M]{\rm Tr}[M^\dagger]$.
Using eqns.~(\ref{qq_cond}) and (\ref{cfl_fac}) we find that the shift 
in the vacuum energy due to the first diagram in Fig.~\ref{fig_4fvac} 
is given by
\be
\label{E_MM}
\Delta {\cal E} = -\frac{3\Delta^2}{4\pi^2} 
 \left\{  \Big( {\rm Tr}[M]\Big)^2 -{\rm Tr}\Big[ M^2\Big]
   \right\}
 + \Big(M\leftrightarrow M^\dagger \Big),
\ee
which agrees with the result of Son and
Stephanov \cite{Son:1999cm}. The two-fermion operators in 
equ.~(\ref{hdet_m}) contribute to the masses of Goldstone bosons
at $O(M^4)$. The corresponding term in the vacuum energy 
was computed in \cite{Bedaque:2001je} 
\be
\label{matchqcd}
\Delta {\cal E} = \frac{m_D^2}{8p_F^2} 
 {\rm Tr}\left[(MM^\dagger)(M^\dagger M)-(MM^\dagger)^2\right],
\ee
where $m_D^2=(21-8\log(2))p_F^2/(36\pi^2)$ is (up to a factor
of $g^2$) the Debye mass in the CFL phase. There are other
contributions to the vacuum energy at $O(M^4)$, but these terms
are suppressed by additional powers of $1/p_F$. 

\section{Anti-Particle Gap Contribution}
\label{sec_agap}

 Several authors have suggested that the calculation 
of the mass shift in the vacuum energy, and consequently
the masses of Goldstone bosons, requires the knowledge of the 
``anti-gap'', that is the gap parameter for the $\psi_-$ excitation 
\cite{Hong:2000ei,Manuel:2000wm,Rho:2000ww,Beane:2000ms,Hong:2000ng}.
As we shall see below, the anti-particle gap contribution to the 
vacuum energy is $\Delta{\cal E}\sim g^{-1} \Delta\overline{\Delta}
M^2$. We shall also see that the natural magnitude of the anti-particle 
gap is $\overline{\Delta}\simeq\Delta$. In this case, the 
anti-particle gap contribution to $\Delta{\cal E}$ 
dominates over the results we obtained in the previous
section. This in itself may not pose a problem, but we
shall also see that a straightforward calculation of the 
anti-particle gap in the microscopic theory does not 
seem to produce a gauge invariant result \cite{Schafer:1999jg}. 

 In order to study the problem in more detail we consider 
the QCD lagrangian in the presence of gap parameters for 
both the $\psi_+$ and $\psi_-$ excitations, 
\bea 
\label{agaps}
{\cal L} &=& \psi_{+,L}^\dagger (iv\cdot D)\psi_{+,L}
 -  \psi_{+,L}^\dagger (i\vec{\alpha}_\perp\cdot\vec{D})\psi_{-,L}
 -  \psi_{-,L}^\dagger (i\vec{\alpha}_\perp\cdot\vec{D})\psi_{+,L}
 +  \psi_{-,L}^\dagger (2p_F+i\tilde{v}\cdot D) \psi_{-,L} \\
 & & \mbox{} + \frac{1}{2} \left(
    \Delta^{AB} \psi^A_{+,L} C\psi^B_{+,L} 
     + \tilde{\Delta}^{AB}_1 \psi^A_{-,L} C\gamma_0\psi^B_{+,R}
     + \tilde{\Delta}^{AB}_2 \psi^A_{+,R} C\gamma_0\psi^B_{-,L}
    \right. \nonumber \\
 & & \left. \hspace{0.7cm}\mbox{}
     + \overline{\Delta}^{AB} \psi^A_{-,L} C\psi^B_{-,L}
     + \Big( h.c. \Big) \right) \nonumber  \\
 & & \mbox{}
     - \psi_{-,L}^\dagger \gamma_0 M \psi_{+,R}
     - \psi_{+,L}^\dagger \gamma_0 M \psi_{-,R}
     + \Big( L\leftrightarrow R, M\leftrightarrow M^\dagger \Big),
     \nonumber 
\eea
where $\tilde{v}=(1,-\vec{v})$. Here, we have included all spin 0 
gaps that can be constructed from $\psi_\pm$. In the CFL phase the 
particle gap has the structure $\Delta^{AB}={\bf \Delta}^{AB}\Delta$ 
with ${\bf \Delta}^{AB}=\delta^{AB}C^A$ and $C^A=(2,-1)$ for $A=
(0,1\ldots 8)$. We shall see that the anti-particle gap has the 
same structure. We will determine the color-flavor structure of 
the mixed particle-anti-particle gap $\tilde{\Delta}$ below. 
In the case of massless quarks, $M=0$, the mixed gap $\tilde{\Delta}$
vanishes in the weak coupling approximation 
\cite{Bailin:1984bm,Schafer:1999na,Pisarski:1999av}.
Since we are interested in mass corrections to the vacuum energy 
density, $\tilde{\Delta}$ cannot be neglected. It is straightforward 
to integrate out $\psi_-$ at tree level. We find
\bea
{\cal L} &=& \psi_{+,L}^\dagger (iv\cdot D)\psi_{+,L}
 - \frac{1}{2p_F}\psi_{+,L} (\Dslash_\perp)^2\psi_{+,L}
 - \frac{1}{2p_F} \psi_{+,L}^\dagger (M M^\dagger) \psi_{+,L} \\
 & & \mbox{} + \frac{1}{2} \Bigg(
    \Delta^{AB} \psi^A_{+,L} C\psi^B_{+,L} 
   - \frac{1}{2p_F} (M^T\tilde{\Delta}_1-\tilde{\Delta}_2M)^{AB}
               \psi^A_{+,R} C\psi^B_{+,R}
        \nonumber \\
 & &  \mbox{} \hspace{0.6cm}
 - \frac{1}{4p_F^2} (M^T\overline{\Delta}M)^{AB} 
      \psi^A_{+,R} C\psi^B_{+,R} 
     + \Big( h.c. \Big)  \Bigg)
    + \Big( L\leftrightarrow R,M\leftrightarrow M^\dagger\Big), 
 \nonumber  
\eea
where $M^{AB}=\frac{1}{2}{\rm Tr}[\lambda^A M (\lambda^B)^T]$ is 
the quark mass matrix $\delta^{ab}M_{ij}$ in the CFL basis. 
We can now determine the contribution of $\overline{\Delta}$
and $\tilde{\Delta}$ to the mass terms in the vacuum energy. 
Computing the diagrams in Fig.~\ref{fig_agapvac} we find
\bea 
\label{agap_vac}
\Delta{\cal E} &=&- \frac{3\pi^2}{2\sqrt{2}g}
     \left(\frac{p_F^2}{2\pi^2}\right)
   \left\{ \frac{1}{(2p_F)^2} 
        {\rm Tr}\Big(M^T\overline{\Delta} M \Delta\Big) 
   + \frac{1}{2p_F}
        {\rm Tr}\Big(M^T\tilde{\Delta}_1\Delta
                    -M\Delta\tilde{\Delta}_2 \Big)
     \right\}   \\
 & & \mbox{} \hspace{3cm}
 + \Big(M\leftrightarrow M^\dagger\Big). \nonumber 
\eea
We can analyze the anti-particle gap contribution in more detail. 
Assuming that the anti-particle gap $\overline{\Delta}$
has the same color-flavor structure as the particle gap 
$\Delta$ we find
\be 
(M^T\overline{\Delta}M)^{AB}
 = -\frac{\overline{\Delta}}{2}
 \left\{ {\rm Tr}\left(M\lambda^A M\lambda^B \right) 
      - {\rm Tr}\left(M\lambda^A\right) 
        {\rm Tr}\left(M\lambda^B\right) \right\}.
\ee
This allows us to write the anti-particle gap contribution 
to the vacuum energy as
\be
\Delta{\cal E} = -\frac{3}{8\sqrt{2}g}\Delta\overline{\Delta} 
   \left\{  \Big( {\rm Tr}[M]\Big)^2 -{\rm Tr}\Big[ M^2\Big]
   \right\} . 
\ee
This result agrees, up to a factor of $1/2$,  with the result 
of Beane et al. \cite{Beane:2000ms}. The factor $1/2$ discrepancy 
is due to the fact that Beane et al.~compute loop integrals 
under the assumption that the gap is not a function of energy
or momentum.

\section{Anti-Particle gap parameters}
\label{sec_ds_agap}

 After we have determined the contribution of the 
anti-particle gap parameters $\overline{\Delta}$ and
$\tilde{\Delta}$ to the mass terms in the vacuum energy 
we shall now try to compute the gap parameters to 
leading order in perturbation theory. This calculation
is performed in the microscopic theory, but we will use
kinematic simplifications that are similar to the 
approximations that are built into the effective 
theory. In principle, computing the anti-particle gap 
parameters requires solving a coupled set of gap equations
for $\Delta$, $\overline{\Delta}$ and $\tilde{\Delta}$.  
However, both $\overline{\Delta}$ and $\tilde{\Delta}$ are 
induced gaps. This means that even though $\overline{\Delta}$ 
and $\tilde{\Delta}$ are not necessarily small, the correction 
to the particle gap $\Delta$ due to the fact that the 
anti-particle gaps are non-zero is suppressed by $1/p_F$. 
As a consequence, only the particle gap $\Delta$ has to be 
determined self-consistently. Both $\overline{\Delta}$ and 
$\tilde{\Delta}$ can be calculated perturbatively. 

 In the last section we saw that $\overline{\Delta}$ 
contributes to ${\cal E}$ at $O(M^2)$ whereas $\tilde{\Delta}$
contributes at $O(M)$. However, $\tilde{\Delta}$ itself is of 
order $(M/p_F)\Delta$. Therefore, both gap parameters effectively 
contribute to ${\cal E}$ at the same in order in $M$. In order to 
collect all contributions to the vacuum energy at $O(M^2)$, we 
shall compute $\Delta$ to $O(M^2)$, $\tilde{\Delta}$ to $O(M)$, 
and $\overline{\Delta}$ to $O(1)$. In practice this is most easily 
achieved by considering the Dyson-Schwinger equation 
\cite{Bailin:1984bm}
\be
\label{ds}
 \Sigma(k) = -ig^2 \int \frac{d^4q}{(2\pi)^4}
  \Gamma_\mu^a S(q)\Gamma_\nu^b D^{ab}_{\mu\nu}(q-k).
\ee
and expanding the quark propagator $S$ in powers of $M$,
\be
 S = S_0 + S_0 {\cal M} S_0 + S_0 {\cal M} S_0 
                 {\cal M} S_0 + \ldots 
\hspace{0.3cm} .
\ee 
Here, $\Sigma(k)$ is the proper self energy in the 
Nambu-Gorkov formalism, $\Gamma^a_\mu$ is the quark-gluon 
vertex and $D^{ab}_{\mu\nu}(q-k)$ is the gluon propagator. To 
leading order in the perturbative expansion, we can use
the free vertex
\be
\label{vert_0}
\Gamma^a_\mu = \left(\begin{array}{cc}
 \gamma_\mu\frac{\lambda^a}{2} & 0 \\
 0 & -\gamma_\mu\left(\frac{\lambda^a}{2}\right)^T \end{array}\right).
\ee
To first order in $1/p_F$, the Nambu-Gorkov propagator 
$S_0$ is given by
\be
 S_0(q) = \left(\begin{array}{cc}
 \frac{q_0+\epsilon_q}{D_+(q)} \gamma_0 \Lambda^-_q 
  + \frac{1}{2p_F} \gamma_0 \Lambda^+_q                 & 
 \frac{\Delta}{D_+(q)} \Lambda^+_q                     \\
 \frac{\Delta}{D_+(q)} \Lambda^-_q                      & 
 \frac{q_0-\epsilon_q}{D_+(q)} \gamma_0 \Lambda^+_q
   - \frac{1}{2p_F} \gamma_0 \Lambda^-_q 
\end{array}\right). 
\ee
Here, $\Lambda^\pm_q=\frac{1}{2}(1\pm\vec{\alpha}\cdot\hat{q})$, 
$D_+(q)=(\epsilon_q^2+\Delta^2)^{1/2}$ and $\epsilon_q=|\vec{q}|
-p_F$. The quark mass matrix has the structure
\be 
{\cal M} = \left( \begin{array}{cc}
 M P_R + M^\dagger P_L & 0 \\
 0 & M^T P_R + M^* P_L 
 \end{array} \right) ,
\ee
where $P_{L,R}$ are left and right handed projection operators. 
In order to determine the gap parameters we only need
the $S_{21}$ component of the Nambu-Gorkov propagator. To $O(M^2)$ 
we find
\bea 
\label{S12_m2}
S_{21}(q) &=& \frac{\Delta}{D_+(q)}(P_R-P_L)\Lambda^-_q \\
 & & \mbox{} - \frac{1}{2p_F D_+(q)}\gamma_0
       \left( M^T\Delta P_R \Lambda^-_q 
              - \Delta M P_L\Lambda^+_q \right)
     \nonumber \\
 & & \mbox{} 
     - \frac{1}{4p_F^2D_+(q)} M^T\Delta M P_L\Lambda^+_q
     + \frac{q_0+\epsilon_q}{2p_F D_+(q)^2}
                           \Delta MM^\dagger P_R\Lambda^-_q
     + \frac{q_0-\epsilon_q}{2p_F D_+(q)^2} 
                           M^TM^*\Delta P_L\Lambda^-_q
     \nonumber \\[0.2cm]
 & & \mbox{} + \Big(M\leftrightarrow M^\dagger,
    L \leftrightarrow R\Big) + \ldots \nonumber
\eea
In order to solve the gap equation (\ref{ds}) we also have 
to specify the gluon propagator. In a general covariant gauge 
the gluon propagator is given by
\be
\label{D_dec}
 D_{\mu\nu}(q) = \frac{P_{\mu\nu}^T}{q^2-G} 
 + \frac{P_{\mu\nu}^L}{q^2-F} - \xi\frac{q_\mu q_\nu}{q^4}
\ee
where the self energies $D$ and $F$ are functions of $q_0$ 
and $|\vec{q}|$, $\xi$ is a gauge parameter, and the projectors 
$P_{\mu\nu}^{T,L}$ are defined by 
\bea
\label{proj}
 P_{ij}^T &=& \delta_{ij}-\hat{q}_i\hat{q}_j , \hspace{1cm}
 P_{00}^T = P_{0i}^T = 0,    \\
 P_{\mu\nu}^L &=& -g_{\mu\nu}+\frac{q_\mu q_\nu}{q^2}
   -P_{\mu\nu}^T .
\eea
To leading order in the coupling constant $g$, and 
in the limit $q_0<|\vec{q}|\ll p_F$ the  electric and magnetic self 
energies $F,G$ are given by
\be
 F = 2m^2, \hspace{1cm}
 G = \frac{i\pi}{2}m^2\frac{q_0}{|\vec{q}|},
\ee
where $m^2=N_fg^2\mu^2/(4\pi^2)$. We can now compute the gap 
parameters using the methods explained in \cite{Schafer:1999jg}. 
If we only take into account the first term in the anomalous
quark propagator equ.~(\ref{S12_m2}) we get 
\be
\label{gap_p}
\Delta^{AB} = {\bf \Delta}^{AB}\frac{g^2}{12\pi^2} 
 \int dq_0\int dx\,
 \left(\frac{\frac{3}{2}-\frac{1}{2}x}
            {1-x + G/(2p_F^2)}
 +\frac{\frac{1}{2}+\frac{1}{2}x}
            {1-x + F/(2p_F^2)} \right)
 \frac{\Delta(q_0)}{\sqrt{q_0^2+\Delta(q_0)^2}}.
\ee
Here, ${\bf \Delta}^{AB}=\delta^{AB} C^A$ with $C^A=(2,-1)$ for 
$A=(0,1\ldots 8)$ is a constant matrix with the color-flavor
structure of the CFL condensate. $\Delta(q_0)$ is the absolute
magnitude of the CFL gap on the quasi-particle mass shell. To 
leading order the equation (\ref{gap_p}) can be solved by taking 
$x\simeq 1$ in the numerator. This means that we approximate scattering 
amplitudes by their values in the forward direction. The particle
gap on the Fermi surface $\Delta\equiv \Delta(p_0=0)$ is given by
\cite{Son:1999uk,Schafer:1999jg,Hong:2000fh,Pisarski:2000tv,Brown:1999aq}
\be 
\label{gap}
 \Delta = 512\pi^4 (2/3)^{-5/2} b_0'\mu g^{-5}
 \exp\left(-\frac{3\pi^2}{\sqrt{2}g}\right),
\ee 
where $b_0'$ is a constant which is determined by non-Fermi liquid
effects \cite{Brown:1999aq} that are not included in our calculation. 
In the same fashion we can also compute the gap for anti-particles.
Again, we only need the first term in the anomalous quark propagator
equ.~(\ref{S12_m2}). We find
\bea
\label{agap_eq}
\overline{\Delta}^{AB} &=& {\bf \Delta}^{AB}
\frac{g^2}{12\pi^2} \int dq_0\int dx\,
 \left(\frac{\frac{1}{2}+\frac{1}{2}x}
            {1-x + G/(2p_F^2)}
 +\frac{\frac{1}{2}-\frac{1}{2}x}
            {1-x + F/(2p_F^2)} \right. \\
 & & \mbox{}\left. \hspace{3cm}
  + \frac{\xi}{1-x + q_0^2/(2p_F^2)}\right)
 \frac{\Delta(q_0)}{\sqrt{q_0^2+\Delta(q_0)^2}}. \nonumber
\eea
We observe that the anti-particle gap has the same color-flavor
structure as the particle gap. To leading order, we can again
approximate $x\simeq 1$ in the numerator. We note that in this
limit there is no contribution from electric gluons. We also 
find that the gauge parameter $\xi$ does not disappear from 
(\ref{agap_eq}). We obtain
\be
\overline{\Delta}^{AB} = \Delta^{AB} (1+3\xi) + O(g\Delta).
\ee
The next step is to include linear mass terms in the 
propagator (\ref{S12_m2}). These terms mix left and right
handed fermions and contribute to the mixed 
particle-anti-particle gaps $\tilde{\Delta}_{1,2}$.
We find
\bea
\label{mgap_eq}
(\tilde{\Delta}_1)^{AB} &=& -({\bf \Delta}M)^{AB}
\frac{g^2}{12\pi^2(2p_F)} \int dq_0\int dx\,
 \left(\frac{\frac{1}{2}+\frac{1}{2}x}
            {1-x + G/(2p_F^2)}
 -\frac{\frac{1}{2}-\frac{1}{2}x}
            {1-x + F/(2p_F^2)} \right. \\
 & & \mbox{}\left. \hspace{3cm}
  + \frac{\xi}{1-x + q_0^2/(2p_F^2)}\right)
 \frac{\Delta(q_0)}{\sqrt{q_0^2+\Delta(q_0)^2}}, \nonumber
\eea
as well as the analogous equation for $\tilde{\Delta}_2$ 
with $({\bf \Delta}M)\to -(M^T{\bf \Delta})$. In equ.~(\ref{mgap_eq})
we dropped contributions that are proportional to $M^\dagger$
as they give terms in the vacuum energy that are proportional 
to ${\rm Tr}(MM^\dagger)$. We note that the magnetic contribution 
and the gauge parameter term have the same structure as the 
corresponding terms in the equation for the anti-particle gap. 
The only difference is that the relative sign of the contribution
from electric gluons is different. We get 
\bea
(\tilde{\Delta}_1)^{AB} &=& -(\Delta M)^{AB}\;\; \frac{1+3\xi}{2p_F}
         + O(g\Delta), \\
(\tilde{\Delta}_2)^{AB} &=& +(M^T \Delta)^{AB}\frac{1+3\xi}{2p_F}
         + O(g\Delta). 
\eea
Finally we determine the $O(M^2)$ shift in the particle gap. 
Again, we drop the $MM^\dagger$ terms in the anomalous quark
propagator equ.~(\ref{S12_m2}). If we write the particle gap 
as $\Delta^{AB}=\Delta^{AB}_{0}+\Delta^{AB}_2+\ldots$ where 
$\Delta^{AB}_0$ is the gap in the limit $M\to 0$ and $\Delta^{AB}_2$ 
is the $O(M^2)$ shift we obtain
\bea
\label{mmgap_eq}
(\Delta_2)^{AB} &=& -(M^T{\bf \Delta}M)^{AB}
\frac{g^2}{12\pi^2(2p_F)^2} \int dq_0\int dx\,
 \left(\frac{\frac{1}{2}+\frac{1}{2}x}
            {1-x + G/(2p_F^2)}
 +\frac{\frac{1}{2}-\frac{1}{2}x}
            {1-x + F/(2p_F^2)} \right. \\
 & & \mbox{}\left. \hspace{3cm}
  + \frac{\xi}{1-x + q_0^2/(2p_F^2)}\right)
 \frac{\Delta(q_0)}{\sqrt{q_0^2+\Delta(q_0)^2}}, \nonumber
\eea
We note that again, the structure of the right hand side is
the same as in the equation for the anti-particle gap. 
We find
\be
(\Delta_2)^{AB} = -(M^T\Delta M)^{AB} \frac{1+3\xi}{(2p_F)^2}
 + O(g\Delta).
\ee
The gap equations for the anti-particle gaps are schematically
shown in Fig.~\ref{fig_agap}. We can now collect all $O(M^2)$ 
contributions to the vacuum energy density. Using (\ref{agap_vac}) 
and including the $O(M^2)$ shift in the gap we have
\bea 
{\Delta}{\cal E} &=& -\frac{3\pi^2}{2\sqrt{2}g}
 \left(\frac{p_F^2}{2\pi^2}\right)
 \left\{  -{\rm Tr}\Big( \Delta_2 \Delta \Big)
   +\frac{1}{2p_F} {\rm Tr}\Big( M^T\tilde{\Delta}_1\Delta
      - M\Delta\tilde{\Delta}_2 \Big) \right. \\
 & & \mbox{}\hspace{3cm}\left.
   + \frac{1}{(2p_F)^2} {\rm Tr}\Big( M^T\overline{\Delta}
         M\Delta\Big)\right\} 
   + \Big( M\leftrightarrow M^\dagger \Big).\nonumber
\eea
Using the results obtained in this section, we can 
express the vacuum energy in terms of the particle gap only. 
We find
\be
{\Delta}{\cal E} = -\frac{3}{16\sqrt{2}g} (1+3\xi)
 \left\{ {\rm Tr}\Big(M^T\Delta M \Delta \Big)
   - 2 {\rm Tr}\Big(M^T\Delta M \Delta \Big)
   +  {\rm Tr}\Big(M^T\Delta M \Delta \Big) \right\} = 0.
\ee
We observe that, to leading order in $g$, the net 
contribution to the vacuum energy from $\overline{\Delta}$, 
$\tilde{\Delta}_{1,2}$ and $\Delta_2$ vanishes. Individually,
all these terms are non-zero but they depend on the gauge 
parameter and therefore have no physical meaning. We also note
that the leading order results for the anti-particle gap
parameters only involve magnetic gluon exchanges. The 
cancellation that we obtained in this section is automatically 
taken into account in 
the effective theory of section \ref{sec_hdetm} because
of the relation (\ref{q+AMq}). Our results indicate 
that it is not necessary to explicitly include 
anti-particle gaps in the high density effective theory. 
There is no instability in the anti-gap channel, and
therefore no reason to include anti-gaps as variational
parameters. The loop diagrams which determine the anti-particle 
gaps are dominated by quasi-particles in the vicinity of the 
Fermi surface. This means that there is also no reason 
to include anti-gaps as effective operators governed 
by short distance effects in the microscopic theory.
The anti-gap contribution to the vacuum energy
is automatically included in the effective theory 
discussed in section \ref{sec_hdetm}.

 Indeed, at next-to-leading order in $g$ there is a non-vanishing
term in the vacuum energy which arises from the contribution
of electric gluons to the anti-particle gaps. This can be 
seen from the fact that the electric gluon contribution
appears with a different sign in eqns.~(\ref{agap_eq},\ref{mmgap_eq}) 
as compared to equ.~(\ref{mgap_eq}). This means that the 
cancellation observed in the magnetic sector does not occur
in the electric sector. It is precisely this effect that we 
computed in section \ref{sec_hdetm} using an effective 
four-fermion vertex. The fact that this is possible can
be seen from the structure of the gap equations. 
To leading order in $g$, both the numerator and the 
denominator of the electric gluon contribution in 
eqns.~(\ref{agap_eq},\ref{mgap_eq},\ref{mmgap_eq}) 
are proportional to $(1-x)$. As a result, the electric
gluon contribution is effectively point-like. 

\section{CFL Chiral Theory (CFL$\chi$Th)}
\label{sec_CFLchi}

 In this section we shall discuss how to match the vacuum
energy calculated in the high density effective theory to the vacuum
energy in the effective chiral theory for the Goldstone modes in 
the color-flavor-locked phase. Since all the relevant steps 
have been discussed in the literature, we can be very brief 
in our presentation. The leading terms in the effective 
chiral Lagrangian take the form
\bea
\label{l_cheft}
{\cal L}_{eff} &=& \frac{f_\pi^2}{4} {\rm Tr}\left[
 \nabla_0\Sigma\nabla_0\Sigma^\dagger - v_\pi^2
 \partial_i\Sigma\partial_i\Sigma^\dagger \right] 
 +\frac{3f_{\eta'}^2}{4} \left[ 
 \partial_0 V\partial_0 V^*  - v_{\eta'}^2 
 \partial_i V\partial_i V^* \right]\\
 & & \mbox{}
 + \Big[ A_1{\rm Tr}(M\Sigma^\dagger){\rm Tr} (M\Sigma^\dagger)V
          + A_2{\rm Tr}(M\Sigma^\dagger M\Sigma^\dagger)V   
          + A_3{\rm Tr}(M\Sigma^\dagger){\rm Tr} (M^\dagger\Sigma)
         + h.c. \Big]+\ldots . 
 \nonumber 
\eea
Here $\Sigma=\exp(i\phi^a\lambda^a/f_\pi)$ is the octet field, 
$f_\pi$ is the pion decay constant, and $A_{1,2,3}$ are the coefficients 
of the $O(M^2)$ mass terms. We have not displayed any field independent 
$MM^\dagger$ terms. The axial $U(1)_A$ field is $V=\exp(4i\theta)=
\exp(2i\eta'/(\sqrt{6}f_{\eta'}))$. As explained in \cite{Bedaque:2001je} 
the chirality conserving two-fermion operators (\ref{kin}) act like an 
effective chemical potential. In the effective chiral theory they 
appear as gauge fields,  
\be
\label{cov}
\nabla_0\Sigma = \partial_0 \Sigma 
 + i \left(\frac{M M^\dagger}{2p_F}\right)\Sigma
 - i \Sigma\left(\frac{ M^\dagger M}{2p_F}\right) .
\ee
The coefficients $A_{1,2,3}$ have to be matched against the $O(M^2)$ 
result for the vacuum energy (\ref{E_MM}). We find 
\be
 A_1= -A_2 = \frac{3\Delta^2}{4\pi^2}, 
\hspace{1cm} A_3 = 0,
\ee
which agrees with the result of Son and Stephanov. We can now
compute the masses of Goldstone bosons in the CFL phase. 
Bedaque and Sch\"afer argued that the expansion parameter 
in the chiral expansion of the Goldstone boson masses is 
$\delta=m^2/(p_F\Delta)$. The first term in this expansion 
comes from the $O(M^2)$ term in (\ref{l_cheft}), but the 
coefficients $A$ contain the additional small parameter 
$\epsilon=(\Delta/p_F)$. In a combined expansion in 
$\delta$ and $\epsilon$ the $O(\epsilon\delta)$ mass term 
and the $O(\delta^2)$ chemical potential term appear at 
the same order. At this order, the masses of the flavored 
Goldstone bosons are
\bea 
\label{mgb}
 m_{\pi^\pm} &=&  \mp\frac{m_d^2-m_u^2}{2p_F} +
         \left[\frac{4A}{f_\pi^2}(m_u+m_d)m_s\right]^{1/2},\nonumber \\
 m_{K_\pm}   &=&  \mp \frac{m_s^2-m_u^2}{2p_F} + 
         \left[\frac{4A}{f_\pi^2}m_d (m_u+m_s)\right]^{1/2}, \\
 m_{K^0,\bar{K}^0} &=&  \mp \frac{m_s^2-m_d^2}{2p_F} + 
         \left[\frac{4A}{f_\pi^2}m_u (m_d+m_s)\right]^{1/2}.\nonumber
\eea
The masses of neutral mesons are unaffected by the effective 
chemical potential term. The elements of the mass matrix are
\be
\begin{array}{rclrcl}
m^2_{11} &=& \frac{8A}{3f_{\eta'}^2}
  \left[ m_s(m_u+m_d)+m_u m_d \right], &
m^2_{13} &=& -\frac{4\sqrt{2}A}{\sqrt{3}f_{\eta'}f_\pi} 
  (m_u-m_d) m_s,  \\
m^2_{33} &=& \frac{4A}{f^2_\pi}
  m_s(m_u+m_d)  , &
m^2_{18} &=&  \frac{8A}{3\sqrt{2}f_{\eta'}f_\pi}
  \left[ m_s(m_u+m_d)-2m_u m_d \right], \\
m^2_{88} &=& \frac{4A}{3f_\pi^2}
  \left[ m_s(m_u+m_d)+4m_u m_d \right], & 
m^2_{38} &=& -\frac{4A}{\sqrt{3}f_\pi^2} 
  (m_u-m_d)m_s.\\
\end{array}
\ee
The flavor-octet and flavor-singlet decay constants $f_\pi$ 
and $f_{\eta'}$ can be matched against the zero-momentum 
axial-vector current correlation functions in the high 
density effective theory. To leading order in $g$ 
\cite{Son:1999cm,Beane:2000ms,Zarembo:2000pj,Miransky:2001bd} 
\be
f_\pi^2 \,=\, \frac{21-8\log(2)}{18} 
  \left(\frac{p_F^2}{2\pi^2} \right), \hspace{1cm} 
f_{\eta'}^2     \,=\, \frac{3}{4} 
 \left(\frac{p_F^2}{2\pi^2} \right).
\ee
We have identified the mass of a Goldstone mode with
the energy of a $\vec{p}=0$ excitation. We observe that the 
mass of the flavored Goldstone modes can become negative. 
Indeed, as observed in \cite{Bedaque:2001je,Kaplan:2001qk}
this is likely to be the case for physically relevant values
of the quark masses and the baryon density. This instability
corresponds to the onset of meson condensation 
\cite{Bedaque:2001je,Kaplan:2001qk}. In the meson
condensed phase the groundstate is reorganized and 
the masses are determined by small fluctuations around 
the new groundstate. This problem was recently studied
in \cite{Miransky:2001tw,Schafer:2001bq}.

\section{Summary}
\label{sec_sum}

 We have studied the effect of non-zero quark masses in effective 
theories of high density quark matter. We first studied the 
effective theory for quasi-particles in the vicinity of the 
Fermi surface. At leading order in $1/p_F$ there is a chirality
conserving two fermion operator which arises from the mass 
correction to the kinetic energy of a quark in the vicinity 
of the Fermi surface. At next-to-leading order we find a 
chirality violating four-fermion operator which originates
from mass corrections to the quark-quark scattering amplitude. 
We have argued that this scattering amplitude can be matched
against a local operator in the effective theory. 

 In the effective chiral theory for Goldstone modes in the 
CFL phase the chirality conserving two-fermion operators
correspond to an effective chemical potential. This effective
chemical potential shifts the energy of flavored Goldstone 
modes. The chirality violating four-fermion operators correspond
to meson mass terms. The coefficients of the meson mass terms
are suppressed by $\Delta/p_F$. We compute the Goldstone 
boson masses and find agreement with earlier results obtained
by Son and Stephanov \cite{Son:1999cm} and Bedaque and Sch\"afer 
\cite{Bedaque:2001je}. 

 We also studied the contribution of the anti-particle gap to 
the masses of Goldstone bosons. We show that to leading order 
in the coupling constant $g$ the contribution from the 
anti-particle gap cancels against contributions from a
mixed particle-anti-particle gap and the $O(M^2)$ shift 
in the particle gap. Individually, all these terms are 
non-vanishing, but they depend on the choice of gauge and
therefore have no physical meaning.

Acknowledgements: We would like to thank S.~Beane, P.~Bedaque,
T.~Fugleberg, D.~Son, and M.~Stephanov for useful 
discussions. This work was supported in part by US DOE 
grant DE-FG-88ER40388.


\newpage 

\begin{figure}
\begin{center}
\leavevmode
\vspace{1cm}
\epsfxsize=10cm
\epsffile{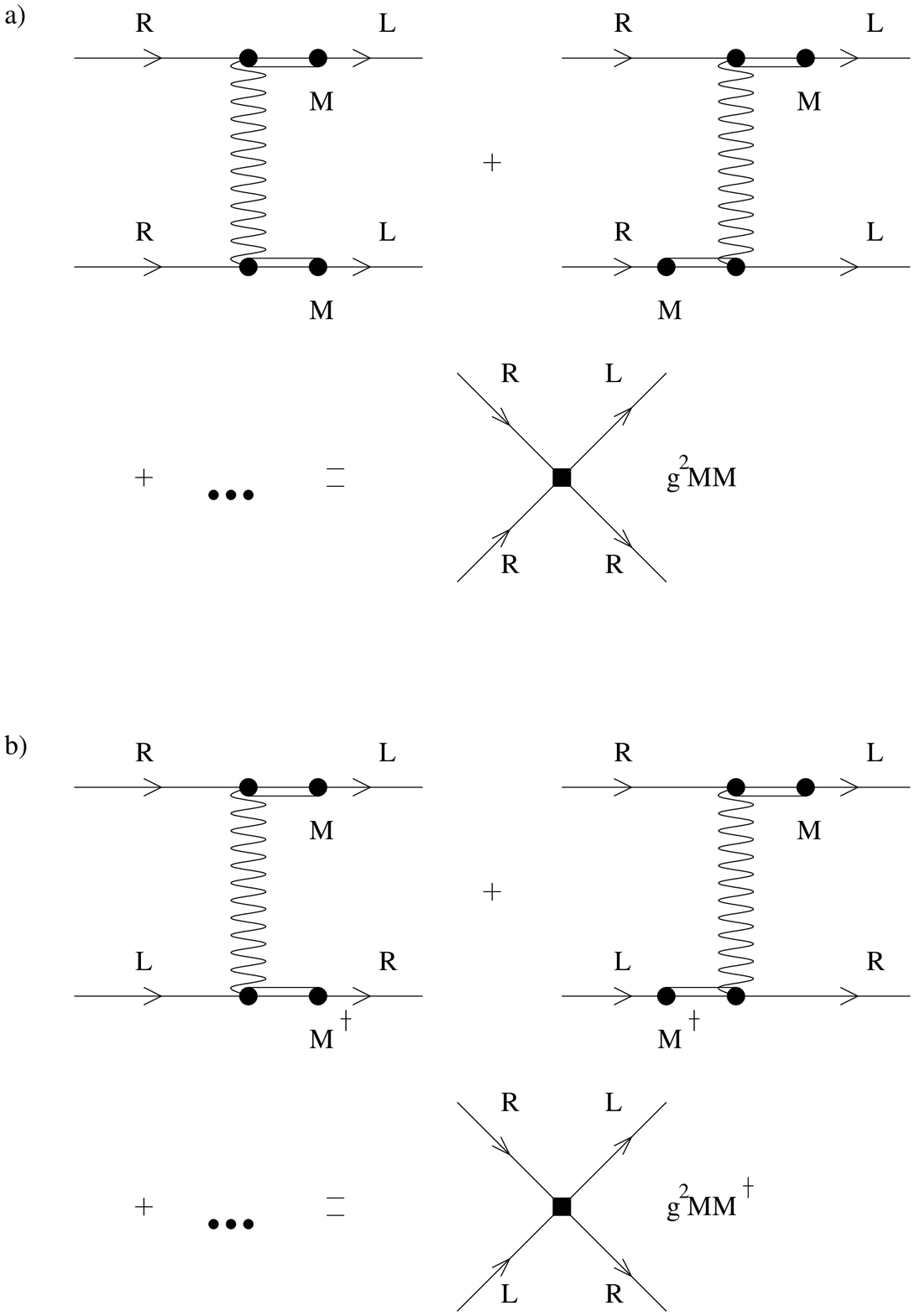}
\end{center}  
\caption{\label{fig_4f}
This figure shows the effective chirality violating four-fermion
vertices. Fig. a) shows the $RR\to LL$ vertex which is proportional
to $MM$ and Fig. b) shows the $RL\to LR$ vertex which is proportional
to $MM^\dagger$. As explained in the text, there are also scattering
amplitudes with two mass insertions on the same fermion line. 
These amplitudes are proportional to $MM^\dagger$, like the 
amplitudes in Fig. b, but they do not violate chirality,
and they cannot be represented as local four-fermion 
operators. }
\end{figure}

\newpage
\begin{figure}
\begin{center}
\leavevmode
\vspace{1cm}
\epsfxsize=12cm
\epsffile{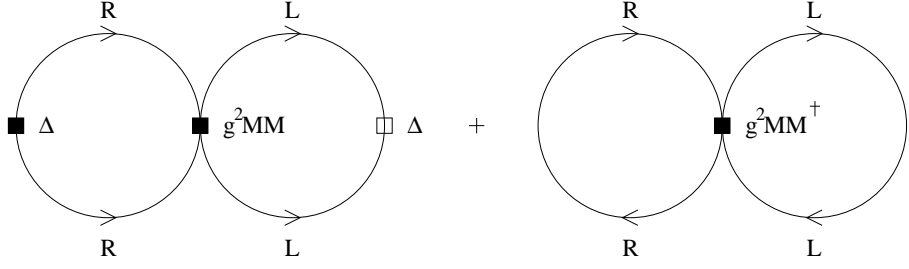}
\end{center}  
\caption{\label{fig_4fvac}
This figure shows the contribution to the vacuum energy
which arises from the four-fermion vertices shown in 
Fig.~\ref{fig_4f}. Note that the second diagram does 
not contribute to the masses of Goldstone bosons. }
\end{figure}

\begin{figure}
\begin{center}
\leavevmode
\vspace{1cm}
\epsfxsize=12cm
\epsffile{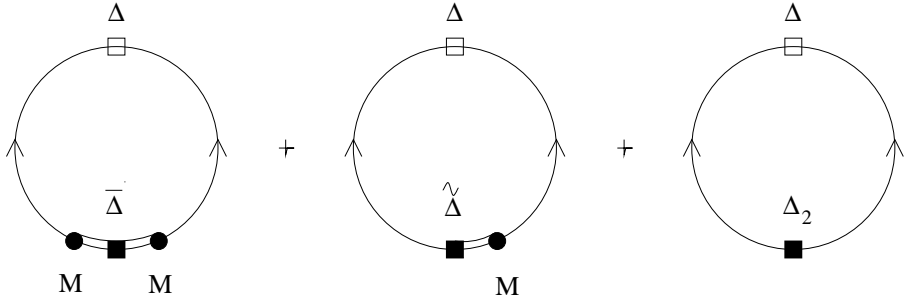}
\end{center}  
\caption{\label{fig_agapvac}
This figure shows the contribution to the vacuum energy
which arises from anti-particle gap, the mixed particle-anti-particle
gap, and the $O(M^2)$ shift shift in the particle gap. As explained
in the text, the sum of all the diagrams shown in this figure vanishes
to leading order in $g$. }
\end{figure}

\newpage
\begin{figure}
\begin{center}
\leavevmode
\vspace{1cm}
\epsfxsize=8cm
\epsffile{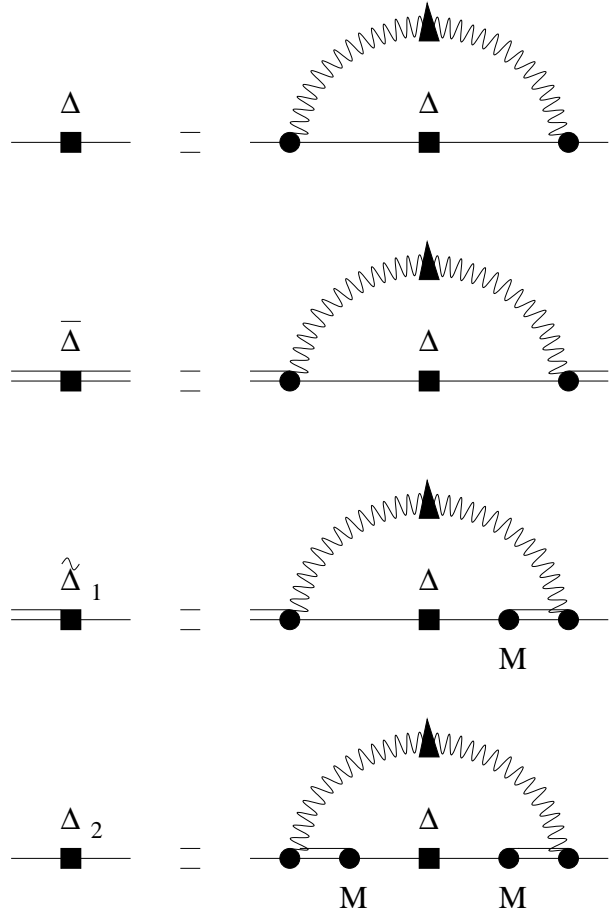}
\end{center}  
\caption{\label{fig_agap}
This figure shows the gap equations for the particle gap 
$\Delta$, the anti-particle gap $\overline{\Delta}$, the 
mixed particle-anti-particle gap $\tilde{\Delta}$, and
the $O(M^2)$ shift in $\Delta$. Solid lines denote $\psi_+$
propagators and double lines denote $\psi_-$ propagators.
The filled circles labeled $M$ denote mass insertions, 
and the filled triangle denotes a gluon self energy insertion. }
\end{figure}

\end{document}